\documentstyle[amsmath,amssymb,multicol,epsfig,prb,aps]{revtex}

\begin{document}
\draft

\title{Gaussian ellipsoid model for confined polymer systems}

\author{Frank Eurich$^1$, Philipp Maass$^{1,2}$ and J\"org
  Baschnagel$^3$}

\address{$^1$Fachbereich Physik, Universit\"at Konstanz, 78457
  Konstanz, Germany}

\address{$^2$Institut f\"ur Physik, Technische Universit\"at Ilmenau,
  98684 Ilmenau, Germany}

\address{$^3$Institut Charles Sadron, 6 rue Boussingault, 67083
  Strasbourg Cedex, France} 

\date{February 22, 2002}

\maketitle

\begin{abstract} 
  Polymer systems in slab geometries are studied on the basis of the
  recently presented Gaussian Ellipsoid Model [J.~Chem.~Phys.
  \textbf{114}, 7655 (2001)]. The potential of the confining walls has
  an exponential shape. For homogeneous systems in thermodynamic
  equilibrium we discuss density, orientation and deformation profiles
  of the polymers close to the walls. For strongly segregated mixtures
  of polymer components $A$ and $B$ equilibrium profiles are studied
  near a planar interface separating $A$ and $B$ rich regions.
  Spinodal decomposition processes of the mixtures in the presence of
  neutral walls show upon strong confinement an increase of the
  lateral size of $A$ and $B$ rich domains and a slowing down of the
  demixing kinetics. These findings are in agreement with predictions
  from time dependent Ginzburg--Landau theory. In the case, where one
  wall periodically favors one of the two mixture components over the
  other, different equilibrium structures emerge and lead to different
  kinetic pathways of spinodal decomposition processes in such systems.
\end{abstract}

\begin{multicols}{2}\narrowtext
  
  \section{Introduction}
\label{sec:introduction}

The tailoring of polymer blends in thin film geometry is a problem of
active current research \cite{Binder:1999} with a wide spectrum of
possible technological applications.  Of particular interest is the
design of suitably patterned structures. Spontaneous phase separation
processes of incompatible polymer blends may be used to translate a
chemical pattern on the surface (as e.g.~produced by the micro-contact
printing technique) into a pattern of varying polymer
compositions.\cite{Boeltau/etal:1998} Another technique developed
recently involves the alignment of polymer melts in inhomogeneous
strong electric
fields.\cite{Schaeffer/etal:2000,Lin/etal:2001,Schaeffer/etal:2001}

Optimization of these processes requires knowledge of the long-time
kinetics of polymeric systems on surfaces.  Unfortunately, direct
numerical investigation of this problem by means of molecular dynamics
simulations is extremely difficult due to the enormous difference
between typical vibrational times ($\sim 1\,{\rm ps}$) and the
characteristic time scales for diffusional processes of the
macromolecules (of the order of seconds). One possible way to tackle 
this problem is to use dynamic self-consistent field
theory,\cite{Reister/etal:2001} which leads to coupled partial
differential equations for the monomer densities of the various
polymer components. Another promising approach was suggested based on
the idea that on semi-macroscopic time and length scales, polymers can
be regarded as soft particles with an ellipsoidal shape
\cite{Murat/Kremer:1998} (a different version of this idea, using soft
spheres instead of ellipsoids, has recently been proposed
\cite{Bolhuis/etal:JCP2001,Bolhuis/etal:PRE2001}). In the work of
Ref.\onlinecite{Murat/Kremer:1998}, a bead spring model with repulsive
Lennard-Jones interactions was used to determine the interaction
parameters of the soft ellipsoids.  Recently, we have set up a similar
ellipsoidal model, where the parameters are derived from the Gaussian
chain model instead.\cite{Eurich/Maass:2001} While the underlying bead spring
model is a more realistic basis for describing polymers, the Gaussian
ellipsoid model (GEM) allows one to deal with effective interaction 
parameters that are independent of the chain length, provided the 
self-interaction of the ellipsoids is taken into account.

In our previous work we have focused on bulk properties of polymer
melts and mixtures, and verified that the model can faithfully
reproduce the basic scaling relations and kinetic properties of
polymeric systems on large time and length scales. In this publication
we extend this study to systems in confined geometries. The
interactions of the ellipsoids with the confining walls are modeled by
a linear coupling of the monomer densities with an external wall
potential. The latter is assumed to have an exponential form with
decay length $l_{w}$, which is numerically very convenient
and allows us to tune the softness of the wall by changing
$l_{w}$. The detailed procedure to account for the wall interactions
is presented in Sec.~\ref{sec:extension}. Some representative results
for homopolymer systems are given in Sec.~\ref{sec:homogeneous}. In
particular, we show density profiles as well as wall induced shape and
orientation effects.  The main focus of this work is, however, to study 
phase separation phenomena in binary mixtures of two polymers $A$ and $B$. 
In Sec.~\ref{sec:demixing} we provide evidence for surface directed
spinodal decomposition waves, when the wall prefers one of the
components $A$ or $B$. In the case of neutral walls, lateral structure
formation is observed in accordance with results derived on the basis
of Ginzburg--Landau type treatments.  Moreover, we quantify the
orientation of the ellipsoids along the phase boundaries.

Finally, in Sec.~\ref{sec:structured} we consider structured systems,
where one wall is considered to be chemically patterned and the
magnitude $\epsilon_{w}$ of the wall potential varies along one wall
direction in a sinusoidal way with periodicity $L_{\parallel}$. This
structuring of the surface induces a periodicity in the polymer melt.
For large film thicknesses periodic polymer structures only occur in
the vicinity of the patterned wall. When the film thickness becomes
comparable to $L_{\parallel}$, the wall induced domain structure
propagates through the film.

 
  \section{Wall Interactions within the Gaussian Ellipsoid Model}
\label{sec:extension}

The Gaussian Ellipsoid Model is described in detail in a previous
publication.~\cite{Eurich/Maass:2001} Therefore, we keep the definition
of the model short and mainly present the necessary extensions to
describe the interaction of the soft particles with the walls.

\subsection{Free Energy Functional}

The free energy functional of an ensemble of $M$ soft particles,
corresponding to polymers with $N$ bonds, is given by an intramolecular
part, which accounts for the possible internal configurations, and an
intermolecular part, which describes the interaction between the
ellipsoids. In the presence of boundaries we add a term which accounts
for the interaction between the particles and the surfaces,
\begin{equation}
F = F_{\text{intra}} + F_{\text{inter}} + F_{\text{wall}}\,.
\label{eq:F}
\end{equation}
The intramolecular part is determined by the probability
$P(\textbf{S})$ for a polymer to have the eigenvalues $\textbf{S}$ of
the radius of gyration tensor $S_{\alpha\beta}$ (in order $S_{x}\ge
S_{y}\ge S_{z}$),
\begin{equation}
  \label{eq:Fintra}
  F_{\text{intra}} = \sum_{i=1}^{M} F_{\text{intra}}^{(i)} =
  -k_{B}T \sum_{i=1}^{M} \ln P(\textbf{S}_{i})\, \text{,}
\end{equation}
where $k_{B}$ is the Boltzmann constant and $T$ is the temperature.
The intermolecular part is given by
\begin{eqnarray}
F_{\text{inter}} &=& \frac{1}{2} \sum_{i=1}^{M} \sum_{j \neq i}
  F_{\text{inter}}^{(ij)}+
  \frac{1}{2} \sum_{i=1}^{M}F_{\text{inter}}^{(ii)}\,,
\label{eq:Fintersum}\\
F_{\text{inter}}^{(ij)}
&=&\hat\epsilon\, b^3\int\hspace*{-0.1cm}d^3 y\,\,
  \varrho_{i}'(\textbf{y})\, \varrho_{j}'(\textbf{y})\,.
\label{eq:Finter}
\end{eqnarray}
where the second term of eq.~(\ref{eq:Fintersum}) accounts for the 
self-interaction of an ellipsoid.  Equation~(\ref{eq:Finter}) shows that
the interaction energy is expressed in terms of the monomer 
densities of ellipsoids $i$ and $j$ (denoted by $\varrho_{i}'(\textbf{y})$ 
and $\varrho_{j}'(\textbf{y})$ in 
the laboratory system \cite{coordinates}).  High monomer concentrations due
to a contraction of a single ellipsoid (self-avoidance) or due to a  strong 
overlap of two ellipsoids are disfavored 
energetically (mutual avoidance).  The parameters
$\hat\epsilon$ and $b^3$ are a ``contact energy'' and a ``contact
volume'', respectively. We denote all reduced energy parameters
without a hat, e.g.  $\epsilon=\hat\epsilon/k_{B}T$. The parameter $b$
sets the length scale of the model, i.e. $b = 1$.

The additional interaction term between the polymers and the walls is
chosen to be of the form
\begin{eqnarray}
  F_{\text{wall}} &=& \sum_{i=1}^{M} F^{(i)}_{\text{wall}} =
  \sum_{i=1}^{M} \int\hspace*{-0.1cm}d^3 y\,\,
  \varrho_{i}'(\textbf{y})\,\, V(z)\,,
  \label{eq:Fwall}\\
  V(z) &=& \hat{\epsilon}_{w} \left[\exp \left(-\frac{z}{2\,l_{w}}\right) + 
        \exp \left(-\frac{L_{z}-z}{2\,l_{w}} \right) \right]\,,
  \label{eq:V}
\end{eqnarray}
where $\hat{\epsilon}_{w}$ is the strength of the monomer--wall
interaction and the wall interaction range $l_{w}$ characterizes the
hardness of the wall. The slab thickness is given by $L_{z}$. Note,
however, that for the special form of the external potential defined
in eq.~(\ref{eq:V}) two parameters $\hat\epsilon_{w}/k_{B}T=1$ and
$L_{z}$ are equivalent to two other parameters
$\hat\epsilon_{w}/k_{B}T=\epsilon_{w}$ and $L_{z}-2l_{w} \ln
\epsilon_{w}$, respectively, assuming all other parameters to be
constant. Without loss of generality we therefore set
$\hat\epsilon_{w}/k_{B}T=1$ in the following and define $L_{z}$ as our
film thickness. This is reasonable, since for $l_{w}\rightarrow 0$ the
system becomes confined by hard walls at positions $z=0$ and
$z=L_{z}$.

An exponential wall potential is frequently used to model a laterally
structureless and purely repulsive wall.\cite{Israelachvili:2000}
Here, it has the advantage, that the overlap between the potential and
the monomer densities, that are of Gaussian form, are analytically
integrable. The formulas are given in appendix
\ref{app:wallinteraction}. By choosing small values for $l_{w}$ one
can model hard substrates, while for large $l_{w}$ one models soft
substrates, that e.g. correspond to polymer films on polymer
brushes.\cite{Grest/Murat:1995} We note that the unperturbed
monomer densities are used in eq.~(\ref{eq:Fwall}).  Hence, we
do not a priori build in layering effects which are typically observed
in simulations of polymer melts close to walls.\cite{Yoon/etal:1995} This 
is in the spirit of the GEM, where the unperturbed chain is used as input 
and all other properties are determined by interaction terms. In summary,
eqs.~(\ref{eq:F}--\ref{eq:V}) define the thermodynamics of the model
for a homogeneous system of ellipsoids in a film geometry.

To generalize the model to binary mixtures of ellipsoids of type $A$
and $B$ in films, we choose an equal interaction between polymers of
the same type $\hat\epsilon_{AA} = \hat\epsilon_{BB} = \hat\epsilon$,
while the interaction between unlike polymers is enhanced by a
mismatch $\delta$, i.e. $\hat\epsilon_{AB} = \hat\epsilon(1+\delta)$.
The same degree of polymerization, i.e.\ $N_{A}=N_{B}=N$, is assumed
for both components. To account for different interactions between the
substrates and the two components, we introduce an additional wall
mismatch parameter $\delta_{w}$, with $-1<\delta_{w}<1$.  The
interaction parameters $\hat\epsilon_{w}(1+\delta_{w})$ and
$\hat\epsilon_{w}(1-\delta_{w})$ refer to the $A$-- and
$B$--components, respectively.  Note that the mismatch parameters
$\delta$ and $\delta_w$ only change the strength of the repulsive
monomer-monomer or monomer-wall interactions.  Phase separation is
thus driven by the disparity of excluded volume interactions, and not
by attraction.

The kinetic properties of the GEM are given by a discrete time
Monte-Carlo algorithm, where three different moves of the ellipsoids
are allowed: Translation of the center of mass, free rotation of the
principal axis and deformation, i.e. a change of the eigenvalues
$\textbf{S}_{i}$. For details we refer the reader to
Ref.~\onlinecite{Eurich/Maass:2001}.
 
\subsection{Parameters and Geometry in Film Simulations}

The simulations in this work were carried out for ellipsoids
corresponding to a chain length of $N=50$. The number of particles is
$M=2000$ in the case of homogeneous systems and $M=4000$ for binary
mixtures. We always choose $\epsilon_{w}=\epsilon=1$.

For binary mixtures a symmetric composition is considered, where the
fraction of the $A$-component is $f_{A}=0.5$.  For given film
thickness $L_{z}$ we determine the lateral dimension $L$ by requiring
a constant ``bulk monomer concentration'' $c=M(N+1)/L_{z}L^{2}=0.85$.
Tables~\ref{tab:geometry.homogeneous} and \ref{tab:geometry.mixture}
list the geometries of the simulated systems.

The homogeneous systems are thermalized several thousand Monte Carlo
steps (MCS) and averaged over at least 10000 MCS. In the case of
binary mixtures, the phase diagrams were determined in the
semi--grand canonical ensemble. Averages are performed over several
thousand MCS.


  \section{Effects of Hard Walls on Homogeneous Systems}
\label{sec:homogeneous}

In this section we present the results for homogeneous ellipsoid
systems bound by two walls. As the focus of this work will be on
binary mixtures in confined geometries, we defer a more detailed
analysis of homogeneous systems in a slab to a forthcoming
publication. 

As a reference length to compare with, we use the unperturbed average
radius of gyration in a bulk system with monomer concentration $\bar
c=0.85$, which is given by, cf.\ Ref.~\onlinecite{Eurich/Maass:2001},
\begin{equation}
  \label{eq:RG.bulk}
  \bar R_{G} = 3.6\,.
\end{equation}

\subsection{Concentration of Monomers and of Centers of Mass}

\begin{figure}[htb]
  \begin{center}
    \epsfig{file=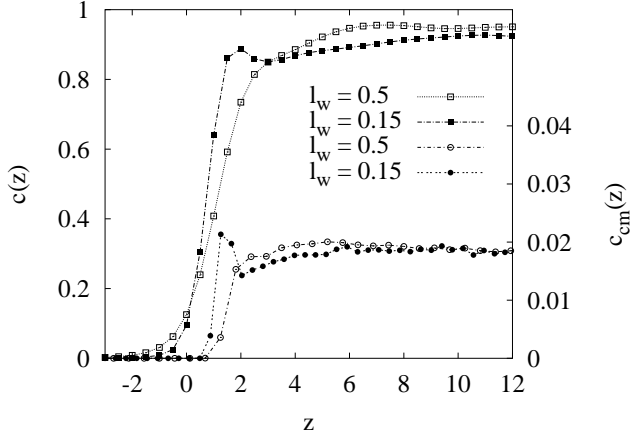,width=1.0\linewidth,angle=0}
  \end{center}
   \caption{Monomer density $c(z)$ (upper curves) and center of mass
     density $c_{\textrm{cm}}(z)$ (lower curves) for wall potentials
     with $l_{w}=0.5$ (open symbols) and $l_{w}=0.15$ (full
     symbols). The film width is $L_{z}=25$.}
   \label{fig:densities.c.ccm}
\end{figure}

The results for the concentration of the monomers
\begin{equation}
  c(z) =  \sum_{i=1}^{M} \int dx\, dy\, \varrho_{i}'(\textbf{y})
\end{equation}
and the concentration of the centers of mass $c_{\textrm{cm}}$ are
shown in Fig.~\ref{fig:densities.c.ccm} for a thick film
($L_{z}\!=\!25$), where bulk-like features are found inside the slab.
Two wall interaction ranges are considered. In the first case,
$l_{w}=0.5$, meaning that $2\, l_{w}$ is equal to the average bond
length of the Gaussian chain molecules, while in the second case the
interaction range is much smaller, $l_{w}=0.15$, modeling a harder
wall. In both cases, the concentration profiles rapidly decay to zero
as $z \rightarrow 0$. However, whereas the profile of the center of
mass, $c_{\textrm{cm}}$, is zero at $z=0$, justifying our definition
of the film thickness, the monomer concentration vanishes only for
$z<0$. This behavior reflects the softness of the wall potential,
which is also responsible for less pronounced layering effects,
typically observed in polymer melts confined between strongly
repulsive walls.\cite{Yoon/etal:1995,Mischler/etal:2001} Here, we find
the signature of these density oscillations only for $l_{w}=0.15$,
whereas the softer wall gives rise to a monomer profile comparable to
that obtained for small bulk concentrations in the hard wall
case.\cite{Yethiraj:1994,Yethiraj/Hall:1990}

\begin{figure}[htb]
  \begin{center}
    \epsfig{file=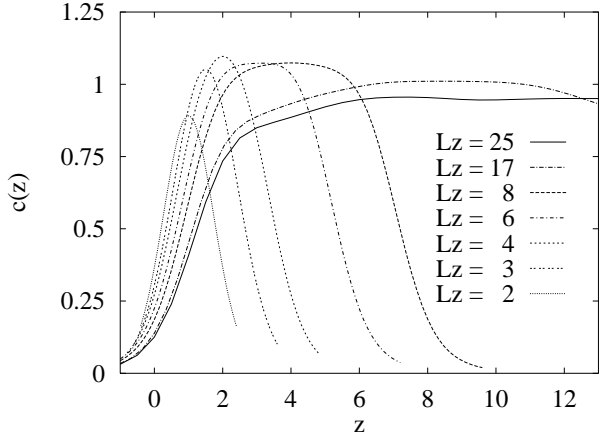,width=1.0\linewidth,angle=0}
  \end{center}
   \caption{Monomer density as a function of distance $z$ from the
     wall for $l_{w}=0.5$ for various film thicknesses $L_{z}$.}
   \label{fig:monomerdensity}
\end{figure}

The monomer concentration profiles for different film thicknesses $L_{z}$
are shown in Fig.~\ref{fig:monomerdensity} for the case $l_{w} = 0.5$.
For the thicker slabs ($L_{z}=17, 25$) a plateau in the profiles is
reached at a distance of about $2\,\bar R_{G}$ from the wall. However,
the plateau value still varies slightly with the system size and is
larger than the bulk value $\bar c=0.85$, that will be obtained for
$L_{z} \rightarrow \infty$. For $L_{z} < 4\, \bar R_{G}$ the films
do not show bulk-like behavior any more. The profile rather runs
through a pronounced maximum $c_{\textrm{max}}$ at $z=L_{z}/2$. 

If $L_{z}$ decreases, starting from $L_z=25$, $c_{\textrm{max}}$ first 
increases, since the reduction of the monomer density close the walls 
becomes more relevant for the overall film behavior, the smaller the
thickness is. It has to be compensated by a higher density in the film
center to keep the density constant. For $L_{z}<4$, however, we observe 
the opposite trend, i.e.\ a decrease of $c_{\textrm{max}}$. This is due to
the fact that the  overlap of the wall potentials becomes important. This
overlap effectively decreases the gradient of the wall potential and thus
allows the monomers to penetrate the wall more deeply.  The increased
monomer density in the wall region entails a reduction of the density 
in the middle of the film in comparison to the thicker films.  The
relative changes of $c_{\textrm{max}}$ are about 20 percent.  In the
case of harder walls ($l_{w}=0.15$) we found a stronger variation and
a largest value $c_{\textrm{max}}=1.36$ is obtained for $L=3$. With
respect to the binary mixtures studied in Sec.~\ref{sec:demixing} we
wanted to keep the influence of these concentration variations small.
We thus chose $l_{w}=0.5$ in the following.  However, very similar 
results could have been obtained for $l_{w}=0.15$.

\subsection{Orientation}
\label{sec:melt.orientation}

In this section we study the orientation of the ellipsoids as a
function of the distance of the center of mass, $z$, from the wall. 

Let us denote by ${\cal R}_{\alpha \beta}$ the matrix, that defines
the orientation of the principal axis with respect to the laboratory
coordinate system. As ${\cal R}_{\alpha \beta}$ consists of the
normalized eigenvectors of the principal axis system,
$\sum_{\beta}{\cal R}_{\alpha \beta}^{2}=1$ holds. In an ensemble of
randomly oriented ellipsoids the average value of all squared matrix
elements is $\langle {\cal R}_{\alpha \beta}^{2} \rangle = 1/3$. Thus
we expect this value for large distances $z$ from the walls. A
parallel orientation of one of the principal axis $\alpha$ to the
laboratory axis $\beta$ corresponds to $\langle {\cal R}_{\alpha
  \beta}^{2} \rangle = 1$, while a perpendicular orientation
corresponds to $\langle {\cal R}_{\alpha \beta}^{2} \rangle = 0$.
Hence ${\cal R}_{\alpha \beta}^{2}$ is a convenient measure to
characterize the orientation. In the following, we denote an average
of some quantity $Q$ over all particles, whose center of mass is at
distance $z$ from the left wall by $\langle Q (z)\rangle$.

\begin{figure}[htb]
  \begin{center}
    \epsfig{file=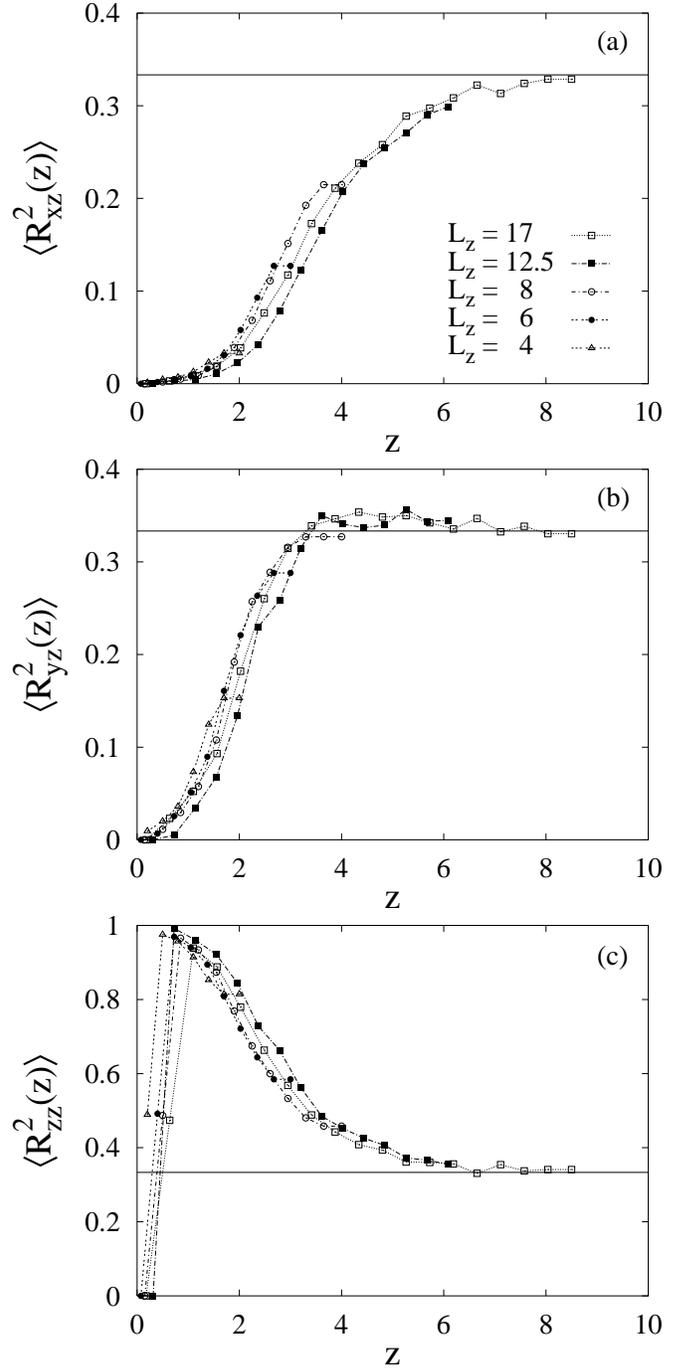,width=1.0\linewidth,angle=0}
  \end{center}
   \caption{Lateral averaged squared matrix elements $\langle {\cal
       R}_{\alpha \beta}^{2}(z) \rangle$ are shown characterizing the
     orientation of the ellipsoids with respect to the $z$-axis of the
     system. The alignment of the principal axis corresponding to the
     largest eigenvalue is given in (a), to the second eigenvalue in
     (b), and to the smallest one in (c). The expected value $1/3$ for
     totally random oriented ellipsoids is marked in as a straight
     line each time. The results are shown for different film widths
     $L_{z}$ and $l_{w} = 0.5$.}
   \label{fig:orientation}
\end{figure}

Figure~\ref{fig:orientation} shows the results for the orientation of
the principal axis with respect to the $z$-axis, i.e.\ perpendicular
to the walls for different film thicknesses. We remind the reader that
the system of principal axes is chosen such that $S_{x}\ge S_{y}\ge
S_{z}$. For thick films ($L_{z}=17$), we find $\langle {\cal
  R}_{\alpha z}^{2}(z) \rangle = 1/3$ far from the walls, as expected.
When approaching the wall the smallest principal axis, $z$, becomes
aligned perpendicular to the wall, $\langle {\cal R}_{zz}^{2}(z) \rangle
\approx 1$, while the other two principal axes become aligned
parallel.\cite{orientation.artefact} This behavior is observed for
all film thicknesses and is also found in other
simulations.\cite{Baschnagel/etal:2000}

The anisotropy in the orientation occurs on a length scale $2 \bar
R_{G}$ at most.  The range is smaller along the $y$--principal
axis than along the $x$--principal axis, since due to $S_{x} >
S_{y}$ the energy cost for a rotation of the $y$--axis in
the direction of the $z$--laboratory axis is smaller than that
of the $x$--principal axis. It is
remarkable that the anisotropy effect shows only very little
dependence on the film thickness, even when $L_{z}$ becomes comparable
to or smaller than $\bar R_{G}$.

\subsection{Deformation}

\begin{figure}[htb]
  \begin{center}
    \epsfig{file=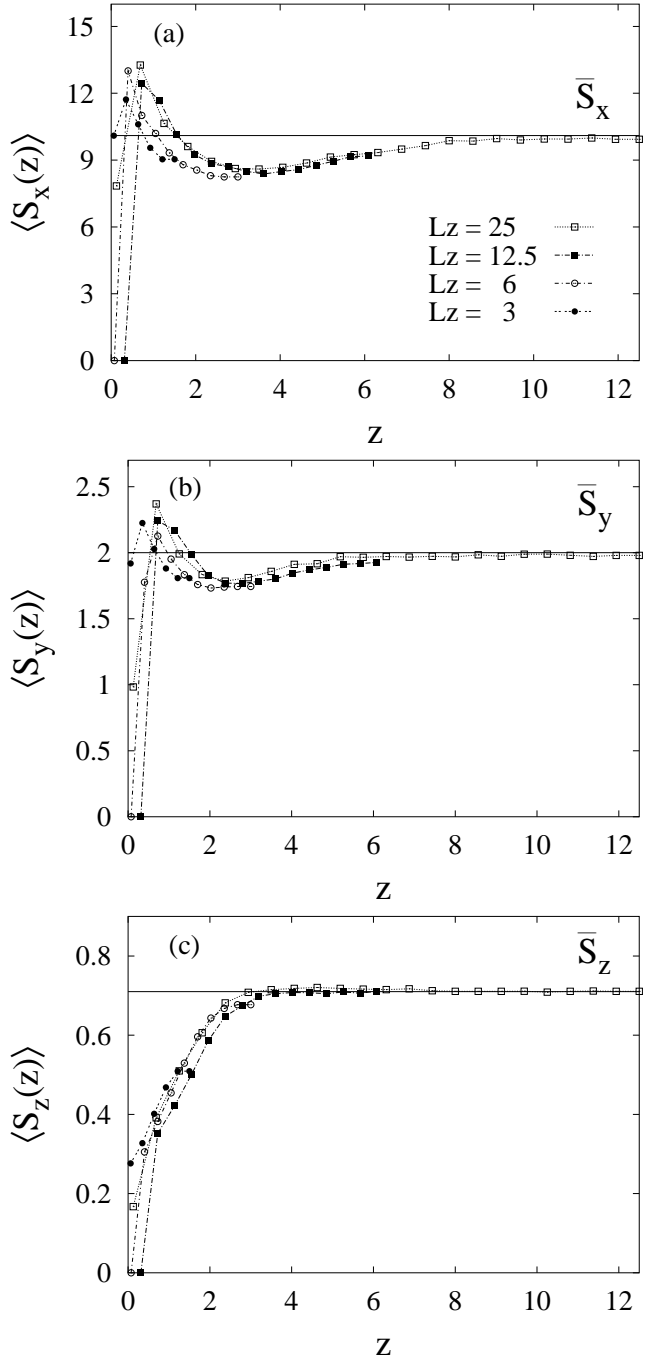,width=1.0\linewidth,angle=0}
  \end{center}
  \caption{Lateral averaged eigenvalues $\langle S_{\alpha}(z)\rangle$
    as a function of the separation $z$ from the wall. The expected
    bulk values from simulations with periodic boundary
    conditions\cite{Eurich/Maass:2001} are shown additionally. The
    results are given for different film widths $L_{z}$ and $l_{w} =
    0.5$.}
   \label{fig:eigenvalues}
\end{figure}

Figure~\ref{fig:eigenvalues} shows how the shape of the ellipsoids is
altered by the presence of the walls.  For large distances, we find that 
the averaged eigenvalues $\langle S_{\alpha}(z) \rangle$ 
are independent of $z$ and coincide with the bulk values.  These values, 
determined from the data of Ref.~\onlinecite{Eurich/Maass:2001}, are 
$\bar S_{x} = 10.1, \bar S_{y}=2.0$ and $\bar S_{z} = 0.71$.

However, the ellipsoids near the walls are distorted.  The size of the
two major axes $x$ and $y$ is enhanced, whereas $\langle S_{z}(z)
\rangle$ is reduced.  When $z$ increases, $\langle S_{z} (z)\rangle$
approaches the bulk value almost monotonously, while $\langle S_{x}(z)
\rangle$ and $\langle S_{y}(z) \rangle$ run through a minimum before
crossing over to the bulk behavior.\cite{orientation.artefact} This
means that the ellipsoids near the surface are strongly oblate.  This
influence of a repulsive wall on the structure of the chains is also
confirmed by other simulations (see Ref.~\onlinecite{Mischler/etal:2001,%
  Baschnagel/etal:2000} and references therein).  As the film
thickness $L_z$ decreases, we find that the maximal values of $\langle
S_{z} (z) \rangle$ become smaller for $L_{z}<2 \bar R_{G}$.

\subsection{Discussion} 

Combining the results of the orientational properties and the
deformation we find that ellipsoids at a distance $z>2 \bar R_{G}$
are virtually unperturbed by the boundaries and exhibit bulk
behavior. On approaching the walls there is a slight deformation of
the two major principal axes, accompanied by an alignment of the
principal axes. This alignment continues monotonously, up to the
point, where the smallest principal axis of the ellipsoids is oriented
perpendicular to the walls. Near the walls, the ellipsoids are
compressed in $z$-direction and elongated with respect to the two
major axes, thus exhibiting an almost two-dimensional behavior.

It is interesting to compare these findings with results
obtained by other authors. Let us mention that the analysis of the
shape of polymers near walls is usually based on ``components'' of the
radius of gyration with respect to the laboratory system. These
components are the square roots of the diagonal elements of the radius
of gyration tensor,
\begin{equation}
  R_{G \alpha} = \sqrt{S_{\alpha\alpha}}\, \textrm{ or } R^{2}_{G
  \parallel} \equiv S_{xx}+S_{yy}\,, R^{2}_{G \perp} \equiv S_{zz}\,. 
\end{equation}
It is clear that 
\begin{equation}
  S_{\alpha\beta} = \sum_{\gamma}{\cal
  R}_{\gamma\alpha}S_{\gamma}{\cal R}_{\gamma\beta}\,,
  \textrm{ esp. } S_{\alpha\alpha} = \sum_{\gamma}{\cal
  R}_{\gamma\alpha}^{2}S_{\gamma}\,.
\end{equation}
This kind of analysis makes it difficult to distinguish between an
orientation and a deformation of the polymer coils. 

When calculating the quantities $\langle R^{2}_{G \parallel} (z)
\rangle $ and $ \langle R^{2}_{G \perp} (z) \rangle$ for our model, we
find reasonable agreement with the results
of other simulations (see Ref.~\onlinecite{Mischler/etal:2001,%
  Baschnagel/etal:2000} and references therein), namely an increase of
$\langle R^{2}_{G \parallel}(z) \rangle$ near the wall and a strong
decrease of $\langle R^{2}_{G \perp}(z) \rangle$. The results for
$\langle R^{2}_{G \parallel}(z) \rangle$ are similar to that for
$\langle S_{x}(z) \rangle$ and $\langle S_{y}(z) \rangle$ in
Fig.~\ref{fig:eigenvalues}, and the results for $\langle R^{2}_{G
  \perp}(z) \rangle$ are similar to $\langle S_{z}(z) \rangle$, except
that the extrema near the walls are more pronounced.

Pai--Panandiker et. al.\cite{Panandiker/etal:1997} examined
orientational effects of the end--to--end vector in a Monte--Carlo
Simulation of dense polymer films. They found an
increasing alignment of the end--to--end vector with reducing
$L_{z}$, but averaged their quantities over the whole film. As there
is a strong correlation between the end--to--end vector and the major
principal axis $x$, their results support the picture presented above. 

The work of Vliet and Brinke\cite{Vliet/Brinke:1990} is dealing with a
principal axis transformation to analyze orientational degrees of
freedom, for a single self--avoiding walk confined between two hard
walls. The main conclusion of this work is that with decreasing film
thickness the principal axis of the polymers are aligned due to the
confinement, long before their shape changes. While their situation is
not directly comparable to the case of a polymer melt between two
plates considered here, our results support the view that the
orientational effect is dominant. This is not unexpected, as a
reorientation of the polymer only constraints on one or two
degrees of freedom, while a deformation affects the whole chain
statistics, therefore affecting of the order of $N$ degrees of
freedom. 


  \section{Demixing Processes in Thin Films}
\label{sec:demixing}

In this section we consider the effect of two neutral ($\delta_{w}=0$)
and planar walls on a binary mixture of ellipsoids. Results will be
discussed for fixed $l_{w} = 1/2$.

\subsection{Dependence of Phase Diagram on Film Thickness} 

For temperatures below a critical temperature $T_{c}$, separation
into two phases, rich in type $A$ and type $B$ polymers, occurs. First,
we determine how $T_{c}$ depends on film thickness $L_{z}$.
Ginzburg--Landau theory\cite{Tang/etal:1993} predicts and
simulations\cite{Rouault/etal:1995} have shown that
\begin{align}
  \label{eq:Lk}
  \frac{T_{c}(\infty)-T_{c}(L_{z})}{T_{c}(\infty)} &\propto
  \frac{1}{L_{z}}\,, \qquad L_{z} \ll N b\,,\\
  \label{eq:Lg}
  \frac{T_{c}(\infty)-T_{c}(L_{z})}{T_{c}(\infty)} &\propto
  \frac{1}{L_{z}^{2}}\,, \qquad L_{z} \gg N b\,.
\end{align}
Equations~(\ref{eq:Lk},\ref{eq:Lg}) hold as long as the chain
conformations are virtually unperturbed, i.e.\ $L_{z} > \bar R_{G}$,
and mean field theory applies.  This requires that the bulk
correlation length $\xi$ (characterizing the interfacial width close
to $T_c$) is smaller than the crossover length\cite{Rouault/etal:1995}
$\xi_{\times} \propto Nb$, which determines the change from
Ising--type to mean field behavior. While for $L_{z} \ll N b$ the
surface interaction is dominant [eq.~(\ref{eq:Lk})], it is the finite
size of the film, which determines the shift of $T_c$ for $L_{z} \gg N
b$ [eq.~(\ref{eq:Lg})].  Since $N=50$ and $2 \leq L_{z} \leq 25$ in
our case, we expect eq.~(\ref{eq:Lk}) to apply, at least for $L_z >
\bar R_{G} = 3.6$ [eq.~(\ref{eq:RG.bulk})] .

\begin{figure}[htb]
  \begin{center}
    \epsfig{file=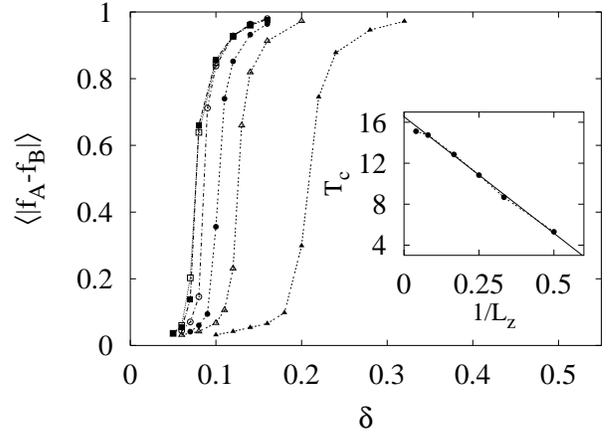,width=1.0\linewidth,angle=0}
  \end{center}
  \caption{Coexistence curves for different film thicknesses
    $L_{z} = 25 (\square), 12.5 (\blacksquare), 6 (\circ), 4
    (\bullet), 3 (\vartriangle), 2 (\blacktriangle)$. The extrapolated
    critical temperatures as a function of $1/L_{z}$ are given in the
    inset.}
   \label{fig:pd}
\end{figure}

For the sake of efficiency the simulations are carried out in a
semi--grand canonical ensemble,\cite{Landau/Binder:2000} i.e.\ the
total number of particles is fixed, while the fractions $f_{A}$ and
$f_{B}$ of $A$ and $B$ polymers can fluctuate. Figure~\ref{fig:pd}
shows the order parameter $\langle|f_{A}-f_{B}|\rangle$ as a function
of the mismatch parameter $\delta$ for various $L_{z}$.

In order to test the expectation that our model obeys
eq.~(\ref{eq:Lk}), we simply estimated $\delta_{c} \propto 1/T_c$ from
an extrapolation of the steepest descent of the coexistence curve to
zero.  No detailed finite size analysis of the lateral dimension $L$
was attempted.  Such an analysis does not seem to be really necessary,
since $T_c(L,L_z)$ monotonously approaches the asymptotic limit
$T_c(L_z)$ for large enough $L$.\cite{Rouault/etal:1995} For the
$L$-values used in this study (see Table~\ref{tab:geometry.mixture}),
inspection of Fig.~5 of Ref.\onlinecite{Rouault/etal:1995} suggests
that qualitatively reliable results should be obtained from our
estimates of $T_c$.  In fact, we see that $T_{c}$ decreases with
$L_{z}$. The inset of Fig.~\ref{fig:pd} shows that
$T_{c}(L_{z})-T_{c}(\infty) \propto - 1/L_{z}$, in agreement with
eq.~(\ref{eq:Lk}).

\subsection{Orientation at Phase Boundaries} 

To study the influence of phase boundaries on the internal degrees of
freedom of the ellipsoids, we prepared a fully phase separated system
with periodic boundary conditions in all directions $L=L_{z}=62.14$.
The $A$--ellipsoids are randomly distributed in the half--space
$z=(0,L_{z}/2)$ and the $B$--ellipsoids in the half--space
$z=(L_{z}/2,L_{z})$.  We chose $\delta = 0.2$, which corresponds to
the strong segregation regime (cf.\ Fig.~\ref{fig:pd}) for thick
films. The system was thermalized for 2000 MCS, a time much larger
than the autocorrelation time of the internal degrees of freedom
(shape and orientation). Moreover, this time is sufficient for an
ellipsoid to diffuse typically over the distance from one interface to
the center of one of the two phases. All averages were recorded over
additional 18000 MCS.

\begin{figure}[htb]
  \begin{center}
    \epsfig{file=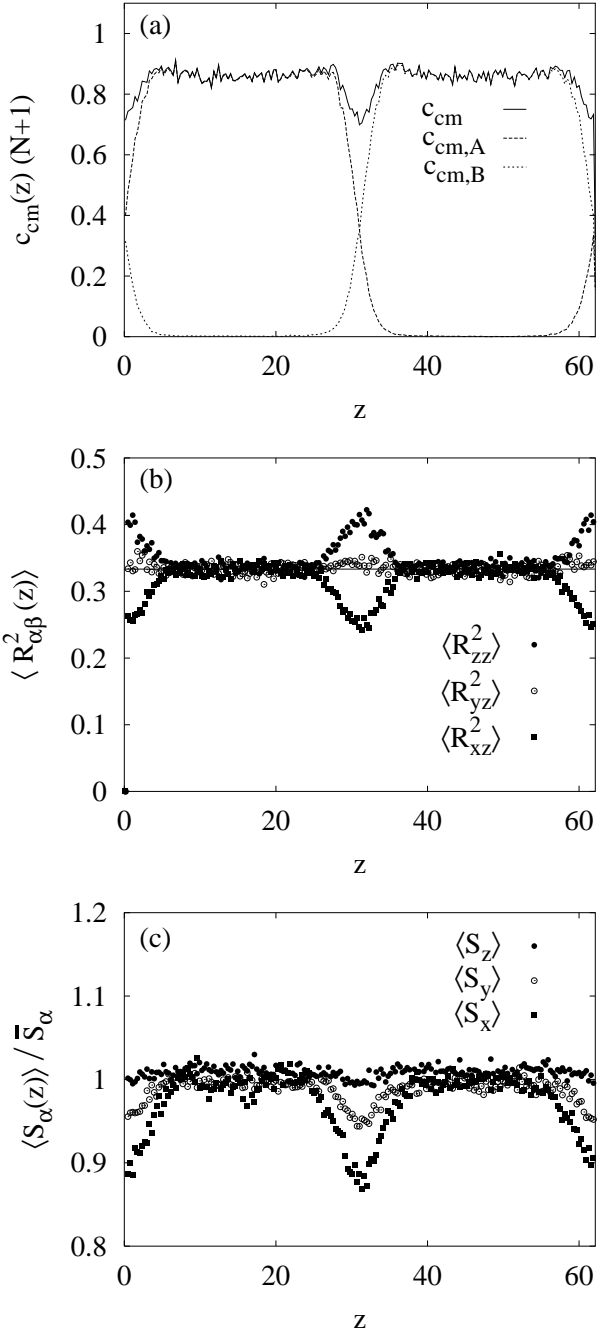,width=1.0\linewidth,angle=0}
  \end{center}
  \caption{Normalized center of mass density (a), squared rotational
    matrix elements (b) and eigenvalues (c) as a function of distance
    $z$ from the interfaces for a separated $A$--$B$ mixture.}
   \label{fig:intface.AB}
\end{figure}

In Fig.~\ref{fig:intface.AB}a the different center of mass densities
for both components $c_{\textrm{cm,}A}$, $c_{\textrm{cm,}B}$, and
$c_{\textrm{cm}}= c_{\textrm{cm,}A}+c_{\textrm{cm,}B}$, normalized by
the chain length ($N=50$) are displayed.  We see that the total
density $c_{\textrm{cm}}$ of the ellipsoids becomes smaller at the
interfaces in order to reduce the interaction between the mixture
components. An additional effect occurs in Fig.~\ref{fig:intface.AB}b,
where the squared matrix elements $\langle {\cal R}_{\alpha
  z}^{2}\rangle$ are shown.  Comparable to the behavior near a wall
(cf.\ Sec.~\ref{sec:melt.orientation}) the ellipsoids align their
$x$--axis (largest principal axis) parallel to the interface, while
the $z$--axis (smallest principal axis) tends to be oriented
perpendicular to the interface.  However, this effect is much less
pronounced than for walls. In particular, there is almost no effect on
the orientation of the $y$--axis.

Finally, Fig.~\ref{fig:intface.AB}c characterizes the deformation of
the ellipsoids via the eigenvalues $S_{\alpha}$ that are normalized by
their bulk values.  Near the interfaces the ellipsoids are compressed
with respect to all axes. The relative compression increases with the
length of the principal axis, meaning that the asphericity becomes
reduced. This effect was also observed in extensive Monte Carlo
simulations of the bond-fluctuation lattice model
\cite{Mueller/etal:1995} and is analogous to the reduction of the
radius of gyration in a strongly segregated mixture, where the
polymers of the minority component shrink to reduce energetically
unfavorable contacts with the the surrounding majority
phase.\cite{Eurich/Maass:2001,Mueller:1998}

\subsection{Spinodal Decomposition Between Neutral Walls}
\label{sec:spinodal.demixing}

Studies based on time--dependent Ginzburg--Landau
theory\cite{Fischer/etal:1998} have shown that for spinodal
decomposition in thin neutral films lateral decomposition waves of a
characteristic wavelength $k_{\parallel,m}$ characterize the domain
pattern at short times, where the concentration inside the $A$ and $B$
rich domains has not yet reached the equilibrium value. These studies
predict that $k_{\parallel,m}$ becomes smaller with decreasing
$L_{z}$, when $L_{z} \lesssim \lambda_{m}$, where $\lambda_{m}$
denotes the bulk demixing length.  Furthermore, if $L_{z}
\lesssim \lambda_{m}$, the growth rates of the decomposition modes
become smaller so that the overall demixing process slows down. These
predictions will be tested here for the GEM.

We consider a system confined between two planar and neutral walls
($\delta_{w}=0$) that is prepared at infinite temperature at a time
$t<0$. All orientations and particle positions are random and the
distribution of the eigenvalues is given by those of free Gaussian
chains. At $t=0$ we quench the system into the spinodal region of the
phase diagram by setting $\epsilon \equiv \hat\epsilon/k_{B}T = 1$
and $\delta = 0.2$ (for $L_{z}=2$ we used $\delta=0.25$). The dynamics
of the system is specified by Monte--Carlo moves of the individual
ellipsoids corresponding to translations, reorientations and
deformations. The detailed procedure is described in
Ref.~\onlinecite{Eurich/Maass:2001}.

To quantify the demixing process of the films, we determine the
lateral intermediate scattering function
\begin{equation}
  \label{eq:I}
  I_{\parallel}(k_{\parallel},z,t) = \langle 
  \rho_{A\parallel}(\textbf{k}_{\parallel},z,t)
  \rho_{A\parallel}(-\textbf{k}_{\parallel},z,t) \rangle - 
  c_{A}^{2} \delta(\textbf{k}_{\parallel})\,, 
\end{equation}
where $\rho_{A\parallel}(\textbf{k}_{\parallel},z,t)$ is the Fourier
transform of the total monomer density $\varrho'_{A}(\textbf{y},t) =
\sum_{i=1}^{M_{A}} \varrho'_{i,A}(\textbf{y},t)$ of the $A$--chains
with respect to the lateral coordinates $x, y$. The power--spectrum
depends on $k_{\parallel}\equiv
\left(k_{x}^{2}+k_{y}^{2}\right)^{1/2}$ only, therefore it is possible
to perform an additional circular average\cite{Kenzler/etal:2001} in
the $k_{x}, k_{y}$--space. An explicit expression for
$\rho_{A\parallel}(\textbf{k}_{\parallel},z,t)$ is given in
appendix~\ref{app:intermediate}.

Figures \ref{fig:intermediate}a-c show
$I_{\parallel}(k_{\parallel},z,t)$ for a slab of thickness $L_{z}=6$
and various fixed times. As one can see from these figures, 
$I_{\parallel}(k_{\parallel},z,t)$ exhibits a maximum that corresponds 
to a characteristic size $2\pi/k_{\parallel,m}$ of the $A$ 
and $B$ rich domains in the lateral direction. With increasing time the
maximum grows and shifts to smaller values of $k_{\parallel}$,
reflecting lateral domain coarsening.

To quantify the speed of this domain coarsening and to examine its
dependence on film thickness $L_{z}$, we consider the first
moment of $I_{\parallel}(k_{\parallel},z,t)$ averaged over $z$,
\cite{Eurich/Maass:2001}
\begin{equation}
  \label{eq:k1}
  k_{1,\parallel}(t) \equiv \frac{1}{L_{z}} \int_{0}^{L_{z}} \!
  dz\, \frac{\int_{0}^{\infty} dk_{\parallel}\,
  k_{\parallel}\, I_{\parallel}(k_{\parallel},z,t)}{\int_{0}^{\infty}
  dk_{\parallel}\, I_{\parallel}(k_{\parallel},z,t)} \,. 
\end{equation}
This quantity is a measure of the average lateral domain size as a
function of time. The results are given in Fig.~\ref{fig:k1} for
different values of $L_{z}$. 

In our system the bulk demixing length $\lambda_{m}$ can be estimated
by $ \lambda_{m} \approx 2\pi / k_{1}(t_{0}) \approx 14$, cf.\ Fig.~15
of Ref.~\onlinecite{Eurich/Maass:2001} ($t_{0}=500\, \textrm{MCS}$).
From this we expect that the effects predicted by time--dependent
Ginzburg--Landau theory should occur for $L_{z}\lesssim 14$. 

Indeed, we find that $k_{1,\parallel}(t_{0})$ determined at
$t_{0}=1000$ MCS shrinks with decreasing $L_{z}$, cf.\ 
Fig.~\ref{fig:k1}. Moreover, in the thinner films the coarsening sets
in later, which is illustrated by the fact that curves at equal
times have a steeper slope for thick films than for thin ones. This
supports the result that the growth rates of the decomposition modes
decrease. We note that these effects are not due to a shift of the
critical temperature $T_{c}$ with $L_{z}$.  We carried out additional
simulations with a mismatch $\delta = 0.3$ for $L_{z}=2,3,4$
and found that the lateral demixing length depends only weakly on the
mismatch parameter at fixed film width.

To summarize this section, our results support the predictions of
Ref.~\onlinecite{Fischer/etal:1998}, i.e.\ the increase of the lateral
demixing length and the slowing down of the demixing kinetics for $L_{z} 
< \lambda_{m}$.  This qualitative agreement with the results of 
Ref.~\onlinecite{Fischer/etal:1998} is obtained without any special wall
potential or tuning of the parameters, supporting the view that the effects
occur quite generally. However, a suppression of spinodal decomposition
below a critical film thickness $L_{c}$, as predicted in
Ref.~\onlinecite{Fischer/etal:1998}, is not found. This is due to the
fact that noise terms in the kinetic equations of the theory have been
neglected.

\subsection{Surface Directed Spinodal Decomposition}

In the case, where one wall preferentially attracts one of the two
components, surface directed spinodal decomposition
waves\cite{Puri/Frisch:1997} (SDW) form when a homogeneous mixture
is quenched into a thermodynamically unstable part of the phase
diagram. However, in thin films the SDW might be ``suppressed'' by the
confinement.\cite{Krausch/etal:1993}

Here, we investigate the case, where both walls attract ellipsoids of
component $B$. Figure~\ref{fig:sdw} shows the emerging monomer
concentrations $c_{B}(z)$, averaged over the lateral dimensions. For
rather large slabs ($L_{z}=50$) symmetric SDW's form, cf.\ 
Fig.~\ref{fig:sdw}a. Their wavelength at short times is given by the
bulk demixing length $\lambda_{m}\approx 14$, and at large times
increases due to domain coarsening. Finally, at $t=100000$ MCS a
$B$--$A$--$B$ stripped pattern has formed in the $z$--direction. In the
case of thin films ($L_{z}=12.5$, cf.\ Fig.~\ref{fig:sdw}b), this kind of
pattern emerges already for short times ($t=1000$ MCS). For
$L_{z}<\lambda_{m}$ the propagation of the SDW is suppressed, and only
a fast segregation of the $B$--components at the walls occur. For even
thinner films ($L_{z}=6$) the two maxima seen in Fig.~\ref{fig:sdw}b,
are less pronounced and merge into one in the case of $L_{z}=3$.


  \section{Structured Surfaces}
\label{sec:structured}

In this section we study polymer mixtures between two walls where one
of the two surfaces is chemically patterned. One part of the patterned
wall attracts component $A$, while the other part attracts component
$B$. The second homogeneous wall is neutral with respect to the two
components. The surface pattern is a periodic arrangement of
equally spaced stripes, cf. Fig.~\ref{fig:structured}, modeled by a
surface potential
\begin{equation}
  \label{eq:Vstructured}
   V(z) = \hat{\epsilon}_{w} \exp \left(-\frac{z}{2\,l_{w}}\right)
     \left[ 1 \pm \delta_{w} \cos\left( \frac{2\pi\,
     y}{L_{\parallel}}\right) \right]\,.
\end{equation}
The positive sign refers to component $A$, the negative to component
$B$. As for homogeneous walls, the overlap integral
[eq.~(\ref{eq:Fwall})] can be calculated analytically, see
App.~\ref{app:wallinteraction}. In the following we consider the case
$l_{w}=0.5$ and $\delta_w=0.5$, where both components are strongly
favored by the respective parts of the wall. Our focus will be on the
influence of the film thickness $L_z$ and of the pattern periodicity
$L_\parallel$ on the demixing process.  

\subsection{Structural Phase Diagram}

From a technological point of view it is important to explore 
regions in $L_z$--$L_\parallel$ space, where the 
pattern of the surface can be translated into a corresponding polymer
structure that pervades the whole
film.\cite{Boeltau/etal:1998,Fukunaga/etal:2000}  

We carried out simulations in a semi--grand canonical ensemble for
different combinations of $L_z$ and $L_\parallel$ summarized in
Tab.~\ref{tab:geo.struct}. As we require $L_y$ to be an integer
multiple of $L_\parallel$ and work with a fixed number of particles
$M=4000$, we adjusted $L_x$ so that the monomer density (or volume)
is constant. The quantities of interest are determined after an
equilibrated stationary domain pattern has evolved.

To characterize the equilibrium patterns we recorded the center of
mass density $c_{\textrm{cm}}$. Two representative results for a slab 
of thickness $L_{z}=6$ are given in Fig.~\ref{fig:struct.y.n}. It shows 
the normalized center of mass density $(N\!+\!1)\, c_{\textrm{cm}}$ as a
function of one laboratory axis ($y$ or $z$), while the density profile
was integrated over the respective
remaining axes. For $L_{\parallel}=25$, Fig.~\ref{fig:struct.y.n}a,
the $A$ and $B$ ellipsoids are separated in a periodic manner due to
the structure of the wall. As one can see from the inset, this
structure propagates through the whole film (``full structure''). By
contrast, for $L_{\parallel}=6$, Fig.~\ref{fig:struct.y.n}b, the
periodic behavior is confined to a small region
near the structured surface (cf.\ the inset). In this case, small
cylindrical caps of one phase are present at the favored parts of the
surface (``partial structure'').  These situations, full and partial
structures, are schematically depicted in Fig.~\ref{fig:structured}.

A systematic study of the emerging structures as a function of $L_{z}$
and $L_{\parallel}$ is shown in Fig.~\ref{fig:struct.pd}. The
($L_{\parallel}$,$L_{z}$)--plane is divided into two regions, where
systems with small $L_{\parallel}$ and large $L_{z}$ are only
partially structured, while systems with large $L_{\parallel}$ and
small $L_{z}$ are fully structured. To explain this behavior, we
associate with the half periods of the cosine term in
eq.~(\ref{eq:Vstructured}) two homogeneous surface regions $\gamma$
and $\delta$, cf.\ Fig.~\ref{fig:structured}.  The contact angles in
these regions are
\begin{equation}
  \cos \Theta_{\gamma} =
  \frac{\sigma_{A\gamma}-\sigma_{B\gamma}}{\sigma_{AB}}\qquad
  \cos \Theta_{\delta} =
  \frac{\sigma_{A\delta}-\sigma_{B\delta}}{\sigma_{AB}}\,.
\end{equation}
Near the $\gamma$--$\delta$ interface the contact angle $\Theta$ can
assume any value between the two limiting cases $\Theta_{\gamma}$ and
$\Theta_{\delta}$.\cite{Lenz/Lipowski:1998} Requiring without loss of
generality that $c_{A} \leq c_{B}$, the contact angle is restricted to
$0 \leq \Theta \leq \pi/2$. A simple consideration of the area of the
interfaces in both cases (a) and (b) given in
Fig.~\ref{fig:structured} yields a difference of the free energy
\begin{equation}
  \Delta F = \sigma_{AB} L_{x} L_{y} \left( 2 L_{z} - \frac{\Theta}{2
  \sin \Theta} L_{\parallel} \right)\,.
\end{equation}
The transition between both structures occurs, if $\Delta F=0$,
resulting in
\begin{equation}
  \label{eq:LzLp}
  \frac{L_{z}}{L_{\parallel}} = \frac{\Theta}{4 \sin \Theta}\,.
\end{equation}
The crossover regime between the two structures in the
$L_z$--$L_\parallel$ diagram is between the two lines with slope $1/4$
(for $\Theta \rightarrow 0$) and slope $\pi/8$ (for $\Theta
\rightarrow \pi/2$). Both lines are marked in Fig.~\ref{fig:struct.pd}
for comparison, showing agreement with the simulations.

This result is also confirmed by experiments of an $A/C$ demixing
process in thin films.\cite{Fukunaga/etal:2000} In the experiment, the
patterned substrate was realized by physisorbing the $B$--component of
an $ABC$ triblock copolymer microphase separated brush. This leads to
a checkerboard like $A/C$ surface structure, instead of the striped
structure considered above. However, the different geometry yields
only a different constant in the righthand side of
eq.~(\ref{eq:LzLp}).

\subsection{Pattern Directed Spinodal Decomposition}

In this section we consider the kinetics of a system similar to the
one described in the previous section, but for fixed fraction
$f_A=0.5$. The initial conditions are as in
Sec.~\ref{sec:spinodal.demixing}.

The intermediate scattering function
\begin{equation}
  I_y(k_y,z,t) = \int_{0}^{\infty} \! d k_{x}\,
  I_{\parallel}(\textbf{k}_{\parallel},z,t)\,,
\end{equation}
with $I_{\parallel}(\textbf{k}_{\parallel},z,t)$ given in
eq.~(\ref{eq:I}) characterizes the domain formation with respect to
the pattern--direction.

Figure~\ref{fig:spin.struct.frozen} shows $I_y(k_y,z,t)$ for a thin
slab of thickness $L_{z}=6$ and pattern periodicity
$L_{\parallel}=25$, corresponding to a full structure in equilibrium,
at two times. At early times ($t=1000$ MCS, see
Fig.~\ref{fig:spin.struct.frozen}a), a dominant structure arises near
the patterned substrate on the length scale $k^{-1}_{y} =
L_{\parallel}/2\pi$. This structure propagates in $z$--direction until
it reaches the opposite wall ($t=10000$ MCS, see
Fig.~\ref{fig:spin.struct.frozen}b). At later times, this structure is
stable, as one would expect from Fig.~\ref{fig:struct.pd}. We
determined no further change in $I_{y}(k_{y},z,t)$ until $t=30000$
MCS.

Figure~\ref{fig:spin.struct.coarse} shows $I_y(k_y,z,t)$ for a film of
thickness $L_{z}=25$ and pattern periodicity $L_{\parallel}=12.5$,
corresponding to a partial structure in equilibrium. In this system the
pattern-induced demixing is much faster than the lateral spinodal
decomposition. The pattern-induced structure for small values of $z$
leads to a much larger corresponding peak in $I_y(k_y,z,t)$,
Fig.~\ref{fig:spin.struct.coarse}a, than the disordered lateral domain
pattern emerging for larger $z$.  The ordered periodic structure
propagates into the film over a finite distance only ($z\approx 6$, see
Fig.~\ref{fig:spin.struct.coarse}b). For later times ($t=30000$ MCS),
the amplitude of the lateral decomposition waves is comparable to the
amplitude of the pattern-induced waves. However, the peak in
$I_y(k_y,z,t)$ occurs at smaller values of $k_{y}$. Thus, the periodic
domain structure is stable near the patterned wall, but further away
the domain pattern coarsens.

Such a behavior, i.e.\ a rapid formation of a patterned equilibrium
structure in thin films, but a lateral domain coarsening afterwards in
thick films, has also been observed in numerical treatments of the
Cahn--Hilliard equation with appropriate boundary
conditions,\cite{Karim/etal:1998,Kielhorn/Muthukumar:1999} and in
experiments.\cite{Karim/etal:1998}

In both cases considered so far the periodicity $L_{\parallel}$ was
comparable to or larger than the bulk demixing length $\lambda_{m}$.
The question arises of what happens for $L_{\parallel}\ll \lambda_{m}$.
In this case, we expect that the patterned surface induces structures,
which are too small to drive spinodal decomposition because the
interfacial energy of such structures is higher than the free energy
gained by decomposition.

To investigate the interplay between the two length scales
$L_{\parallel}$ and $\lambda_{m}$ we consider a slab of thickness
$L_{z}=12.5$ and vary $L_{\parallel}$.\cite{slab.demixing.length}
Figure~\ref{fig:spin.struct.interplay} shows corresponding lateral
structure factors $I_y(k_y,z,t)$ for $t=1000$ MCS and for
$L_{\parallel}=6,12.5,$ and $25$. In the case of $L_{\parallel}=6 \ll
\lambda_{m}$, Fig.~\ref{fig:spin.struct.interplay}a, the peaks
corresponding to the surface induces structure and the lateral modes
are comparable. The demixing process is not dominated by the surface-induced
pattern.  Also, the induced structure develops much more
slowly compared to the case of $L_{\parallel}=12.5 \simeq
\lambda_{m}$, shown in Fig.~\ref{fig:spin.struct.interplay}b. In this
case, the peak corresponding to the ordered structure is a factor of
about 10 higher, and the lateral demixing modes are thus almost
invisible. For $L_{\parallel}=25 \gg \lambda_{m}$ a double peak
structure emerges.  This can be interpreted in terms of a pattern
directed spinodal decomposition wave, i.e.\ if one looked at the
concentration fluctuations $\delta c_{A}(\textbf{y}) =
\varrho_{A}'(\textbf{y})- c_{A}$ in the $yz$--plane and labeled positive
and negative regions of $\delta c_{A}(\textbf{y})$ black and white
respectively, one would see a checkerboard-like
pattern.\cite{Kielhorn/Muthukumar:1999}

To see more directly how in the last case such a pattern propagates
into the film, we plotted in Fig.~\ref{fig:pattern.directed.sd} the
intermediate scattering function $I_{y}(2\pi/L_{\parallel},z,t)$ as a
function of $z$ for various times $t$ for a system with $L_{z}=25$ and
$L_{\parallel}=50$. This picture describes how the surface-induced
periodic domain patterns grow in perpendicular direction. Since our
film thickness is limited, we did not try to quantify the propagation
of this kind of kinetics further. For very large film thicknesses we
would expect the usual Lifshitz--Slyozov domain growth to occur at
large times.

        
  \section{Summary}
\label{sec:summary}

The Gaussian Ellipsoid Model \cite{Eurich/Maass:2001} (GEM) is based
on the fact that the most probable shape of a polymer is not
spherical, but
ellipsoidal.\cite{Solc/Stockmayer:1971,Aronovitz/Nelson:1986,Janszen/etal:1996}
This ellipsoid resembles a flattened American football, the
eigenvalues $\textbf{S}$ of the three principal axes being all
different.  In the GEM, the eigenvalues and their distribution are
calculated for a Gaussian chain.  Thus, the model should faithfully
represent the large scale chain structure in a $\Theta$-solvent or in
a melt.  In order to extend the model to dilute and semi-dilute
solutions in good solvents excluded volume interactions have to be
added.  This is achieved by determining the local monomer density in
the Gaussian ellipsoid.  A strong overlap of the density in the
ellipsoid or between different ellipsoids entails a large energy
penalty which mimics the self- and mutual excluded volume interaction
of soft polymer coils.

In the bulk, the model reproduces well established properties on large
length and time scales.\cite{Eurich/Maass:2001} For homopolymers, it
yields the correct crossover scaling from dilute to semi-dilute
solutions and the correlation hole in the melt.  For symmetric
homopolymer mixtures, one (approximately) finds a linear dependence of
the critical temperature on chain length, a contraction of the chain
size of the minority component, and a Lifshitz-Slyozov behavior for
the late stages of spinodal decomposition.

These results prompt us to test the model if interfaces are present.
These interfaces are either self-generated, as it is the case in the
strong-segregation regime of binary polymer mixtures, or imposed
externally by the introduction of two repulsive walls on opposite
sides of the simulation box.  The main results of this extension may
be summarized as follows: 

(1) In the vicinity of a repulsive wall the
ellipsoids can no longer adopt all possible orientations.  They become
aligned and distorted.  The two longer axes are stretched and oriented
parallel to the wall, whereas the short axis shrinks and orients
perpendicular to it.  These wall-induced perturbations of the chain's
structure cross over to the isotropic bulk behavior for large film
thicknesses (i.e., $L_z > 4 \bar R_G$).  These results parallel those
obtained from less coarse-grained simulation models
(Refs.~\onlinecite{Mischler/etal:2001,Baschnagel/etal:2000} and
references therein).  

(2) Chains close to the interface in a strongly
segregated polymer mixture behave similarly.  In order to reduce
unfavorable interactions the largest axis of the ellipsoids orients
parallel to the interface and shrinks in size so as to increase the
number of favorable self-contacts inside the chain.  These results
qualitatively agree with those obtained in a detailed Monte Carlo
study of the phase behavior of polymer blends by the bond-fluctuation
model.\cite{Mueller/etal:1995}

(3) If a binary mixture of ellipsoids between two neutral walls
undergoes spinodal decomposition, a lateral demixing length
$\lambda_{\parallel,m}$ is found, which increases for films thinner 
than the bulk demixing length ($L_{z}<\lambda_{m}$). These results 
are in agreement with time--dependent Ginzburg--Landau
theory.\cite{Fischer/etal:1998} Experimental studies of this effect
have, to the best of our knowledge, not been performed yet. However,
both the present simulations and theory lead to similar results and 
suggest that such studies should be promising.

(4) In the case where one of the two walls is structured, two
different equilibrium patterns are found. Either the structure
propagates through the whole film, leading to a striped demixing pattern,
or the surface-induced structure is confined to the
vicinity of the wall. Both types of pattern are separated by a
straight line in the $L_{z}$--$L_{\parallel}$ space, as can be
explained by a consideration of surface energies only. Different
equilibrium patterns lead to different kinetic pathways for the
spinodal decomposition in such a film. In the case of an equilibrium
pattern with full structure, pattern directed spinodal decomposition
is the dominant process. The periodic ordered pattern freezes after
reaching the opposite wall. On the other hand, the penetration depth of
the pattern directed spinodal wave remains finite and lateral domain
coarsening takes place in the late stages of decomposition. In the
case of small pattern periodicity ($L_{\parallel}<\lambda_{m}$),
pattern directed spinodal decomposition is less pronounced, as the
pattern induces unfavorable structures with too large interfaces.

In summary, the presented results suggest that the GEM is a promising model
to study large length and time scales not only in the bulk, but also in the 
presence of interfaces.  To test the model further we envisage to analyze
the dependence of the structure of a thin polymer film on chain length. 
Several simulations \cite{Wang/Binder:1991,Bitsanis/Hadziioannou:1990}
have already been devoted to this problem, with which we can compare our
results.  Furthermore, recent experiments \cite{Kraus/etal:2000,%
Jones/etal:2001} investigated the dependence of the radius of gyration on 
film thickness and obtained diverging results.  We hope to contribute to this
current discussion by the simulations envisaged.
  
\acknowledgements

We should like to thank W.~Dieterich, B.~D\"unweg, R.~Everaers and
K.~Kremer for interesting and sti\-mu\-la\-ting discussions. Financial
support by the Deut\-sche Forschungsgemeinschaft and the Minist\`ere de 
l'Edu\-ca\-tion Nationale, de la Recherche et de la Technologie in the 
framework of the European Graduate College ``Soft Condensed Matter'' 
(Konstanz--Strasbourg--Grenoble) is gratefully acknowledged.

 
  \begin{appendix}

\section{Calculation of Ellipsoid-Wall Interaction}
\label{app:wallinteraction}

The integrals in eq.~(\ref{eq:Fwall}) with a potential of the form
eq.~(\ref{eq:V}) and monomer densities, that are essentially sums and
products of Gaussian functions, cf.\ eqs.~(25a-c) in
Ref.~\onlinecite{Eurich/Maass:2001}, can be cast into the more general
form already treated in the appendix of
Ref.~\onlinecite{Eurich/Maass:2001}. The six resulting integrals of
eq.~(\ref{eq:Fwall}) are of the form
\begin{equation}
  \label{eq:wall.overlap.integral}
  \int_{-\infty}^{+\infty}\hspace*{-0.2cm} d^3y\,\,
  \chi(\textbf{y})\,\psi(y_{3})\,,
\end{equation}
where the functions $\chi(\textbf{y})$ and $\psi(y_{3})$ are  
\begin{eqnarray}
  \label{eq:chipsidef}
  \chi(\textbf{y})&=&\frac{1}{(2\pi)^{3/2} \prod_{\alpha=1}^3
    \sigma_{\alpha}} \exp\left(-\frac{1}{2}\sum_{\alpha} 
    \left(\frac{x_{\alpha}(\textbf{y})}{\sigma_{\alpha}}
    \right)^{2} \right)\,,\nonumber\\[0.2cm]
  &&  \hspace*{-1cm}
  x_{\alpha}(\textbf{y})=\sum_{\beta} {\cal R}_{\alpha
    \beta} (y_{\beta}-w_{\beta}) \,, \hspace*{0.3cm}
  \sigma_{\alpha} = R_{\alpha}\sigma_{\alpha}\quad \text{and}
  \nonumber\\[0.2cm] 
  &&  \hspace*{-1cm}
  \psi(y_{3}) = \exp \left(\pm \frac{y_{3}}{2\,l_{w}}\right)
    \label{eq:psi}\,. 
\end{eqnarray}
We adopted the notation and definitions of
Ref.~\onlinecite{Eurich/Maass:2001}. An analogous calculation yields
the same result for the integral as in eqs.~(C12-C14) of
Ref.~\onlinecite{Eurich/Maass:2001}, except for the values of $\cal
A_{\alpha \beta}, B_{\alpha}$ and $C$, that in the present case read
\begin{eqnarray}
  {\cal A}_{\alpha \beta}&=& {\cal L}_{\alpha \beta}\,,\\ 
  B_{\alpha}  &=& \mp \frac{\delta_{\alpha 3}}{l_{w}} - 2 \sum_{\beta} 
  {\cal L}_{\beta \alpha} w_{\beta} \label{eq:B}\,,\\ 
  C &=& \sum_{\alpha,\beta} {\cal L}_{\alpha \beta} w_{\alpha}
  w_{\beta}\,. 
\end{eqnarray}
The negative sign in front of the Kronecker-delta in eq.~(\ref{eq:B})
corresponds to the positive sign in eq.~(\ref{eq:psi}) and vice versa.

In the case of a patterned substrate, we define
$l_{\parallel}=L_{\parallel}/4\pi$ and write for the first part of the
wall potential 
\begin{equation}
  \label{eq:Vstruc}
  V(y_{3}) = \hat\epsilon_{w} \exp \left(-\frac{y_{3}}{2l_{w}} \right)
  \left[ 1 \pm 
  \delta_{w} \cosh \left(\frac{iy_{2}}{2l_{w}}\right) \right] \,.
\end{equation}
Thus the integration of the respective exponential functions can be
performed as above. Only the form of $B_{\alpha}$ changes, which now
reads
\begin{equation} 
  B_{\alpha} = \frac{\delta_{\alpha 3}}{l_{w}} \mp
  \frac{\delta_{\alpha 2} i}{l_{\parallel}} - 2 \sum_{\beta} 
  {\cal L}_{\beta \alpha} w_{\beta}\,,
\end{equation}
where the negative sign refers to the first imaginary exponential
function in eq.~(\ref{eq:Vstruc}), while the positive sign refers to
the second one, respectively.

We note, that in our implementation the interaction between the walls
and the ellipsoids is never cut off, in contrast to the interactions
between the ellipsoids.\cite{Eurich/Maass:2001}

\section{Calculation of the Lateral Fourier Transform of the Density}
\label{app:intermediate}

\noindent
The calculation of
\begin{equation}
  \rho_{\parallel}(\textbf{k}_{\parallel},y_{3}) = \frac{1}{M(N\!+\!1)}
  \sum_{i=1}^{M} \int_{-\infty}^{+\infty} \hspace{-0.2cm} d^{2}y_{\parallel} 
  \,  e^{-i \textbf{k}_{\parallel}\cdot\textbf{y}_{\parallel}}
  \varrho'_{i}(\textbf{y}_{\parallel},y_{3}) 
\end{equation}
is analogous to the derivation of $\rho(\textbf{k})$ given in appendix
D of Ref.~\onlinecite{Eurich/Maass:2001}. Therefore we only present
the result,
\begin{multline}
  \label{eq:rhopara}
  \rho_{\parallel}(\textbf{k}_{\parallel},y_{3}) = \frac{1}{\sqrt{2\pi}}
  \frac{1}{(2+c_{1})\sigma_{1}\sigma_{2}\sigma_{3}} \frac{1}{M}
  \sum_{i=1}^{M} \frac{1}{\sqrt{S_{1}^{(i)}S_{2}^{(i)}S_{3}^{(i)}}} \\
  \frac{1}{\sqrt{\textrm{det}\,{\cal G}^{(i)}}} \frac{1}{\sqrt{2\,
      {\cal G}_{33}^{(i)^{-1}}}} \exp\left(-\frac{1}{2}
    \sum_{\alpha=1}^{2}\sum_{\beta=1}^{2} k_{\alpha}
    {\cal G}_{\alpha\beta}^{(i)^{-1}} k_{\beta} \right) \times\\
  \exp \left[ \frac{1}{2} \left( \frac{ i \left(y_{3}-r_{3}^{(i)}\right)}{\sqrt{2\,{\cal G}_{33}^{(i)^{-1}}}} - 
      \frac{1}{2\sqrt{2\, {\cal G}_{33}^{(i)^{-1}}}}
      \sum_{\beta=1}^{2} k_{\beta} G_{3\beta}^{(i)}
    \right)^{2} \right] \times \\
  \exp \left(-i\,\sum_{\beta=1}^{2} k_{\beta}r_{\beta}^{(i)} \right)
  \times \left\{ \exp\left(-\frac{1}{2} \frac{\left( {\cal
  R}_{13}^{(i)} \bar{u}_{1} \right)^{2} S_{1}^{(i)}} {2\,{\cal
  G}_{33}^{(i)^{-1}}} \right) \times \right. \\ 
\left. 2\,\cos \left[  \frac{{\cal
  R}_{13} \bar{u}_{1} \sqrt{S_{1}^{(i)}}}{2\,{\cal
  G}_{33}^{(i)^{-1}}} \left( i \left(y_{3}-r_{3}^{(i)} \right)  - \frac{1}{2}
  \sum_{\beta=1}^{2} k_{\beta} G_{3\beta}^{(i)} \right) \right. \right.  \\ 
\left. \left. \phantom{ \frac{{\cal
  R}_{13} \bar{u}_{1} \sqrt{S_{1}^{(i)}}}{2\,{\cal
  G}_{33}^{(i)^{-1}}} } +\sum_{\alpha=1}^{2} k_{\alpha} {\cal
  R}_{1\alpha} \bar{u}_{1} \sqrt{S_{1}^{(i)}} \right]  + c_{1}\right\}\,,
\end{multline}
with
\begin{align}
  \label{eq:appB.G}
  {\cal G}_{\alpha\beta}^{(i)} &\equiv \sum_{\gamma}
  g_{\gamma}^{(i)} {\cal R}_{\gamma\alpha}^{(i)}{\cal
    R}_{\gamma\beta}^{(i)} \\
  \label{eq:appB.G2}
  G_{3\beta}^{(i)} &\equiv {\cal G}_{\beta3}^{(i)^{-1}}\!\! 
  +{\cal G}_{3\beta}^{(i)^{-1}} \text{ and} \\ 
  \label{eq:appB.g}
  g_{\gamma}^{(i)} &\equiv \left(
  \left(\sigma_{1}^{2}S_{1}^{(i)}\right)^{-1},
  \left(\sigma_{2}^{2}S_{2}^{(i)}\right)^{-1},
  \left(\sigma_{3}^{2}S_{3}^{(i)}\right)^{-1} \right)\,.
\end{align}
The quantities ${\cal R}_{\alpha\beta}^{(i)}$ and $r_{\alpha}^{(i)}$
denote the rotation matrix and the center of mass of the $i$'th
particle respectively. The constants
$\sigma_{1},\sigma_{2},\sigma_{3},c_{1}$ and $\bar{u}_{1}$ are given
in Tab.~V of Ref.~\onlinecite{Eurich/Maass:2001}.  Using
eqs.~(\ref{eq:rhopara}-\ref{eq:appB.g}) it is straightforward to
calculate the lateral intermediate scattering function,
eq.~(\ref{eq:I}).
\end{appendix}




\clearpage

\begin{table}[!h]
  \caption{System sizes for homogeneous films.}
  \begin{center}
    \begin{tabular}{lccccc}
      $L_{z}$         & 25     & 17     & 12.5   &  10    &   8    \\
      $L_{x}, L_{y}$  & 69.28  & 84.02  & 97.98  & 109.54 & 122.47 \\
      \hline
      $L_{z}$         &   6    & 4      &   3    &   2    &   1.5  \\
      $L_{x}, L_{y}$  & 141.42 & 173.21 & 200.00 & 244.95 & 282.84 \\
    \end{tabular}
  \end{center}
  \label{tab:geometry.homogeneous}
\end{table}

\begin{table}[!h]
  \caption{System sizes for films of binary mixtures.}
  \begin{center}
    \begin{tabular}{lccccccc}
      $L_{z}$   & 50 & 25 & 12.5 & 6 & 4 & 3 & 2 \\
      $L_{x}, L_{y}$ & 69.28 & 97.98 & 138.56 & 200.00 & 244.95 &
      282.84 & 346.41 \\
    \end{tabular}
  \end{center}
  \label{tab:geometry.mixture}
\end{table}

\begin{table}[!h]
  \caption{System sizes for films of binary mixtures on structured
  surfaces.} 
  \begin{center}
    \begin{tabular}{lccccccc}
      $L_\parallel = 50$: & & & & & \\
      $L_z$ & 50 & 25 & 12.5 & 6 & 4 \\
      $L_y$ & -- & 100 & 150 & 200 & 250 \\
      $L_x$ & -- & 96.26 & 128.34 & 200.53 & 240.64\\
      \hline
      $L_\parallel = 25$: & & & & & \\
      $L_z$ & 50 & 25 & 12.5 & 6 & 4 \\
      $L_y$ & 75 & 100 & 150 & 200 & 250 \\
      $L_x$ & 64.17 & 96.26 & 128.34 & 200.53 & 240.64\\
      \hline
       $L_\parallel = 12.5$: & & & & & \\
      $L_z$ & 50 & 25 & 12.5 & 6 & 4 \\
      $L_y$ & 75 & 100 & 137.5 & 200 & 250 \\
      $L_x$ & 64.17 & 96.26 & 140.01 & 200.53 & 240.64\\
      \hline
      $L_\parallel = 6$: & & & & & \\
      $L_z$ & 50 & 25 & 12.5 & 6 & 4 \\
      $L_y$ & 72 & 102 & 138 & 204 & 246 \\
      $L_x$ & 66.84 & 94.37 & 139.50 & 196.60 & 244.55\\
   \end{tabular}
  \end{center}
  \label{tab:geo.struct}
\end{table}


  

\clearpage
\newpage

\vspace{6cm}

\vspace{1cm}

\vspace{14cm}

\vspace{3cm}

\begin{figure}
  \begin{center}
    \epsfig{file=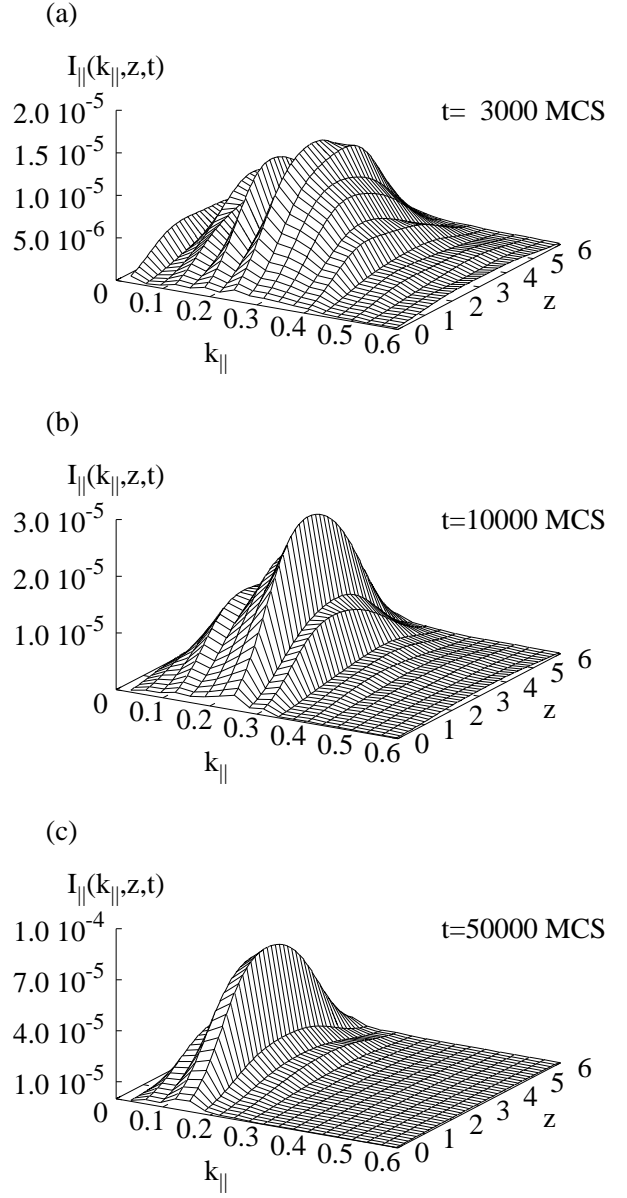,width=1.0\linewidth,angle=0}
  \end{center}
  \caption{Lateral intermediate scattering function
    $I_{\parallel}(k_{\parallel},z,t)$ for a film of thickness
    $L_{z}=6$ shown at various times (a) $t=3000$ MCS, (b) $t=10000$
    MCS and (c) $t=50000$ MCS.}
  \label{fig:intermediate}
\end{figure}

\vspace{3cm}

\begin{figure}
  \begin{center}
    \epsfig{file=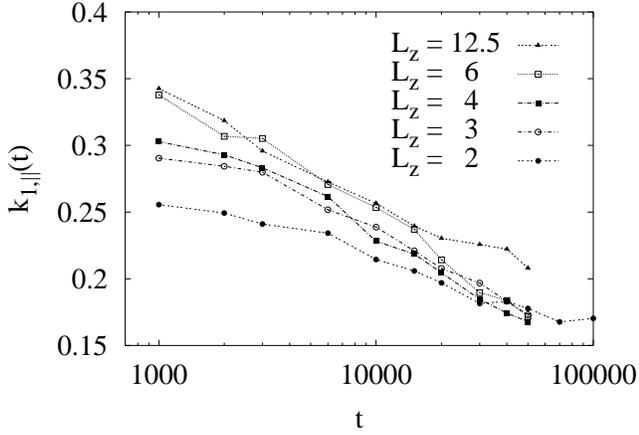,width=1.0\linewidth,angle=0}
  \end{center}
  \caption{Film averaged first moment of the lateral intermediate
    scattering function $k_{1,\parallel}(t)$, eq.~(\ref{eq:k1}), as
    a function of time for different film thicknesses $L_{z}$.}
  \label{fig:k1}
\end{figure}

\begin{figure}
  \begin{center}
    \epsfig{file=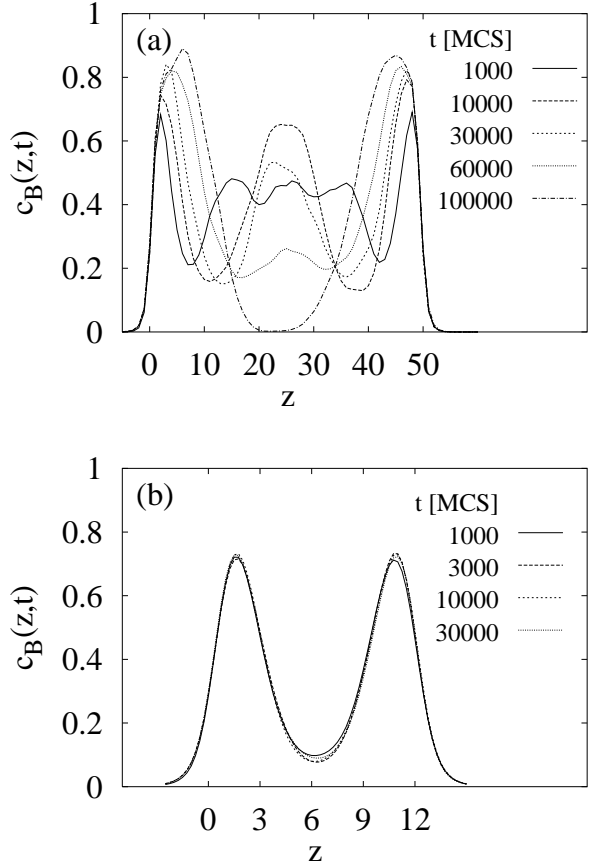,width=1.0\linewidth,angle=0}
  \end{center}
  \caption{Laterally averaged monomer concentration $c_{B}(z)$ of the
    $B$--ellipsoids, for different times, in a film of thickness (a)
    $L_{z}=50$ and (b) $L_{z}=12.5$. The mismatch wall interaction is
    $\delta_{w}=0.5$ for both walls.}
  \label{fig:sdw}
\end{figure}

\begin{figure}
  \begin{center}
    \epsfig{file=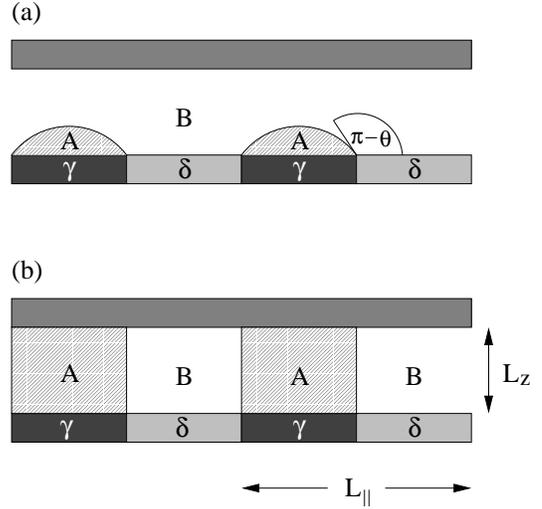,width=0.8\linewidth,angle=0}
  \end{center}
  \caption{Two possible structures on a patterned substrate. Upper
    figure (a): The polymer film is only patterned near the surface
    (``partial structure''). Lower figure (b): The polymer structure
    induced by the surface penetrates the film (``full structure'').}
  \label{fig:structured}
\end{figure}

\begin{figure}
  \begin{center}
    \epsfig{file=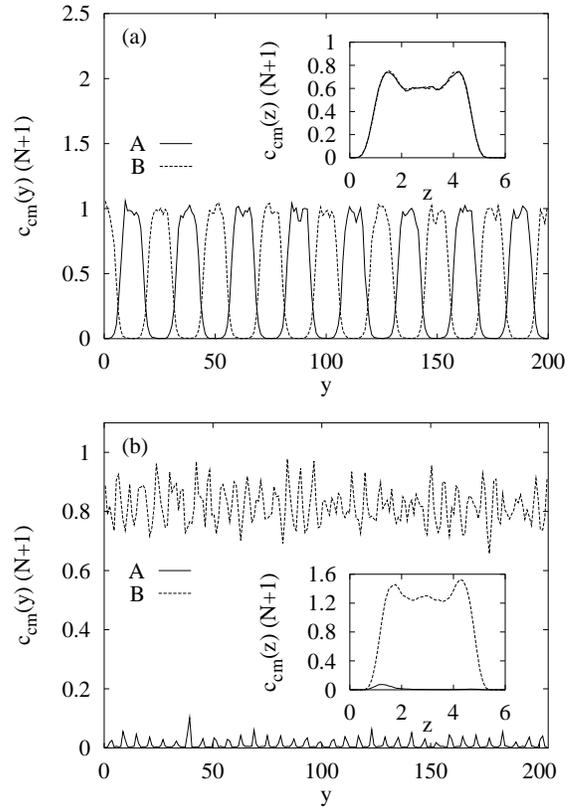,width=0.9\linewidth,angle=0}
  \end{center}
  \caption{Results for the normalized center of mass
    density $c_{\textrm{cm}}$ as a function of $y$ and $z$ (shown in
    the respective inset) in the case of (a) full, and (b) partial
    structure. The film thickness is $L_{z}=6$ in both cases and the
    periodicity of the structure is (a) $L_{\parallel}=25$ and (b)
    $L_{\parallel}=6$.}
  \label{fig:struct.y.n}
\end{figure}

\begin{figure}
  \begin{center}
    \epsfig{file=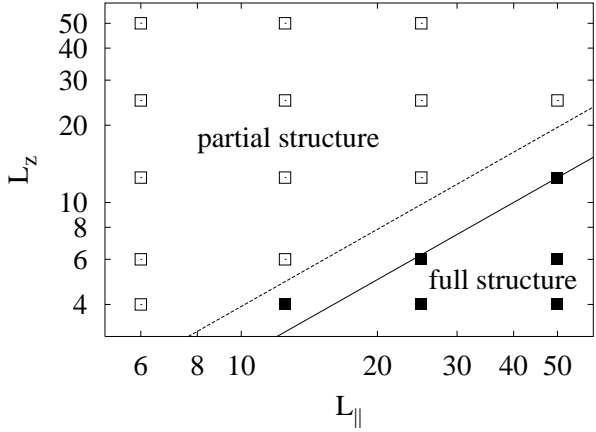,width=1.0\linewidth,angle=0}
  \end{center}
  \caption{Regions in the $L_{z}$--$L_{\parallel}$ plane, where the
    domain pattern in the slab is fully structured ($\blacksquare$)
    and only partially structured ($\square$), in double logarithmic
    representation. The dividing lines are given by the two limiting
    cases of eq.~(\ref{eq:LzLp}), $L_{z} = 1/4\, L_{\parallel}$
    (straight line) and $L_{z} = \pi/8\, L_{\parallel}$ (dashed
    line).}
  \label{fig:struct.pd}
\end{figure}

\begin{figure}
  \begin{center}
    \epsfig{file=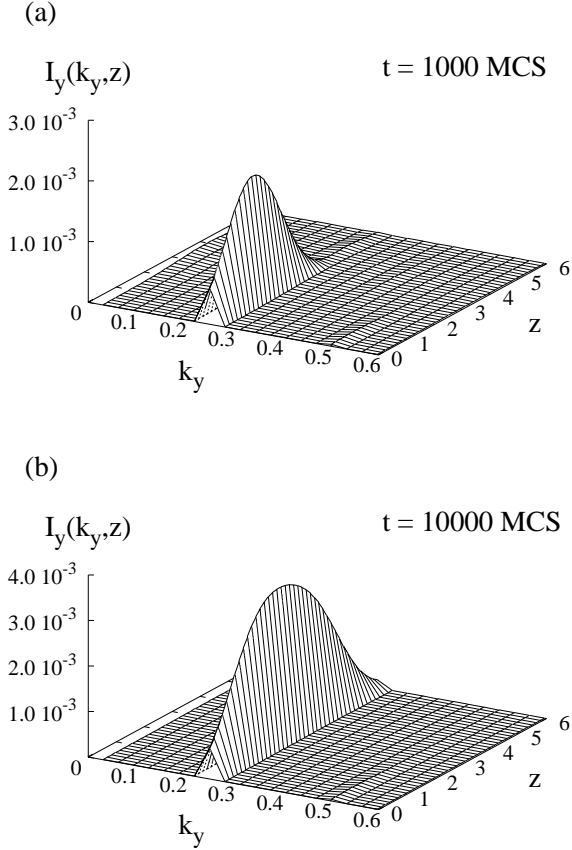,width=1.0\linewidth,angle=0}
  \end{center}
  \caption{Intermediate scattering function $I_{y}(k_{y},z,t)$ for a
    system of $L_{\parallel}=25$ and $L_{z}=6$ at times (a) $t=1000$
    MCS and (b) $t=10000$ MCS. The patterned structure is stable
    throughout the whole film.}
  \label{fig:spin.struct.frozen}
\end{figure}

\begin{figure}
  \begin{center}
    \epsfig{file=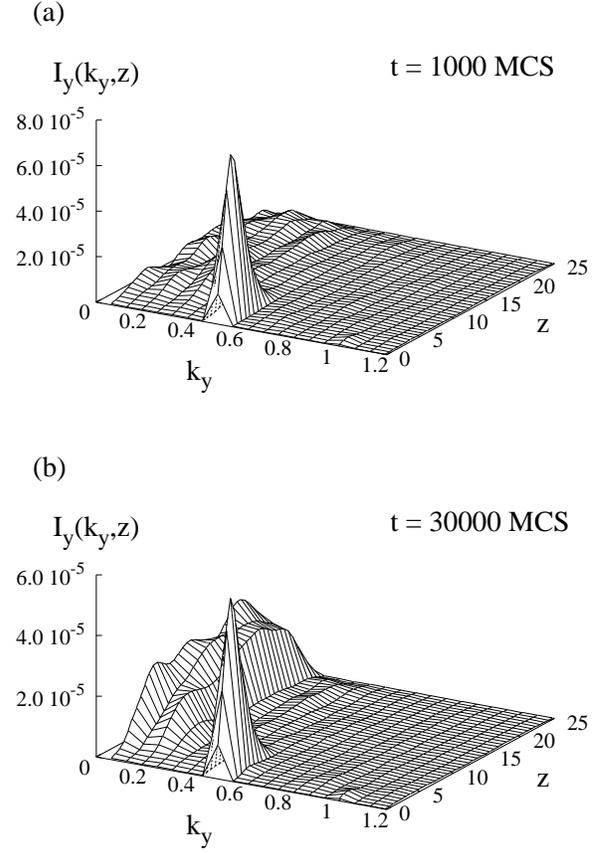,width=1.0\linewidth,angle=0}
  \end{center}
  \caption{Intermediate scattering function $I_{y}(k_{y},z,t)$ for a
    system of $L_{\parallel}=12.5$ and $L_{z}=25$ at times (a) $t=1000$
    MCS and (b) $t=30000$ MCS. Lateral domain coarsening occurs away
    from the vicinity of the patterned wall. Near the wall the
    structure changes only weakly.}
  \label{fig:spin.struct.coarse}
\end{figure}

\vspace{8cm}

\begin{figure}
  \begin{center}
    \epsfig{file=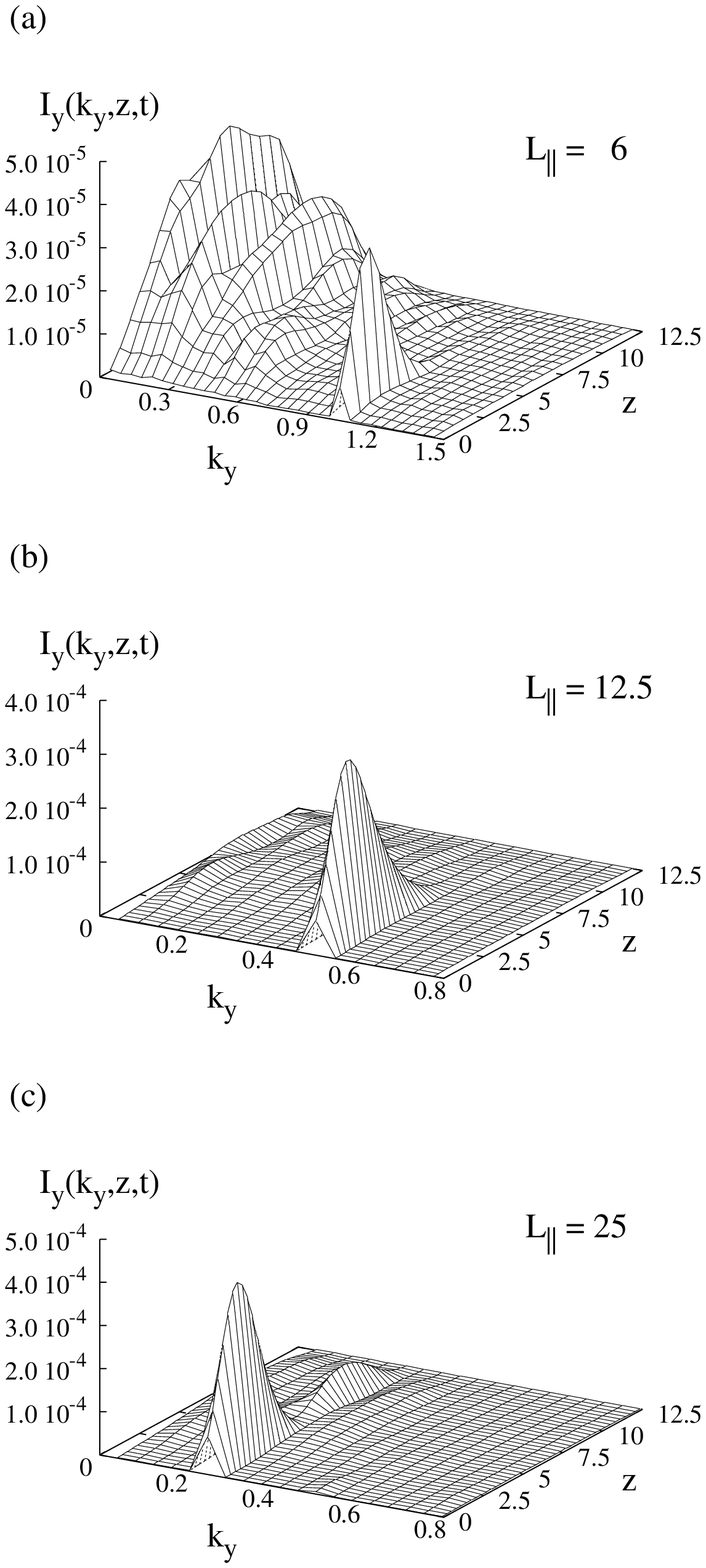,width=1.0\linewidth,angle=0}
  \end{center}
  \caption{Intermediate scattering function $I_{y}(k_{y},z,t)$ for
    early demixing times $t=1000$ MCS and film thickness
    $L_{z}=12.5$. The pattern periodicity is (a) $L_{\parallel}=6$, (b)
    $L_{\parallel}=12.5$ and (c) $L_{\parallel}=25$.}
  \label{fig:spin.struct.interplay}
\end{figure}

\vspace{3cm}

\begin{figure}
  \begin{center}
    \epsfig{file=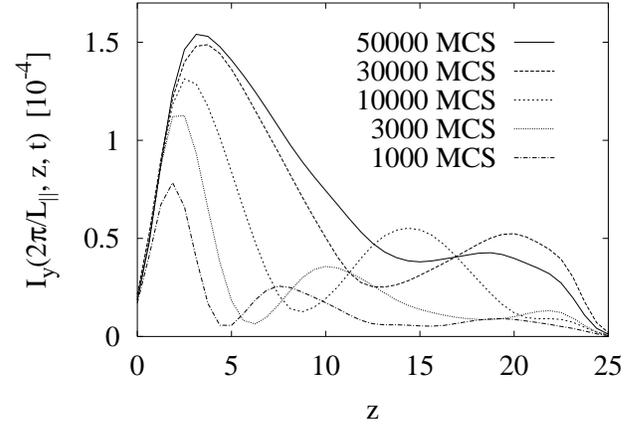,width=1.0\linewidth,angle=0}
  \end{center}
  \caption{Intermediate scattering function
    $I_{y}(2\pi/L_{\parallel},z,t)$ as a function of time. The film
    thickness is $L_{z}=25$ and the pattern periodicity
    $L_{\parallel}=50$.}
  \label{fig:pattern.directed.sd}
\end{figure}

\clearpage


\end{multicols}


\begin{thebibliography}{99}
\bibitem{orientation.artefact} We used the symmetry with respect to
  $L_{z}/2$ in the representation of the data. The non--continuous
  drop to zero at $z\approx 0$ is an artefact of the analysis, because
  the center of mass density goes to zero in this region, cf.\ 
  Fig.~\ref{fig:densities.c.ccm}.
\bibitem{slab.demixing.length} Strictly speaking,
  $\lambda_{\parallel,m}$ is the relevant length scale for lateral
  decomposition in a slab. However, for $L_{z}=12.5$ the demixing
  length $\lambda_{\parallel,m}$ is close to the bulk value
  $\lambda_{m}$, cf.\ Sec.~\ref{sec:spinodal.demixing}.
\end{thebibliography}
\end{document}